\documentclass[a4paper]{jpconf}
\usepackage{graphicx}
\begin{document}
\title{Protons in High  Density Neutron Matter}

\author{Misak M. Sargsian}

\address{Department of Physics, Florida International University, Miami, FL 33199, USA}

\ead{sargsian@fiu.edu}

\begin{abstract}
We discuss the possible implication of the recent predictions of  {\em two new properties}  of 
high momentum distribution of nucleons in asymmetric nuclei for neutron star dynamics.
The {\em first} property  is about the  
approximate scaling relation between proton and neutron high momentum 
distributions weighted by their relative fractions ($x_p$ and $x_n$) in the nucleus. The {\em second}  is the 
existence of inverse proportionality of the high momentum distribution strength  of  protons and neutrons to $x_{p/n}$.
Based on these predictions we model the high momentum distribution functions  for asymmetric nuclei and 
demonstrate that it describes reasonably well the high momentum characteristics  of light nuclei. 
We also extrapolate our results to heavy nuclei as well as infinite nuclear matter
and calculate the relative fractions of protons and neutrons  with momenta above $k_{F}$.
Our results indicate that  for neutron stars  starting at {\em three} nuclear saturation 
densities the protons with  $x_p = {1\over 9}$ will populate mostly the high momentum  tail of 
the momentum distribution while only 2\% of the neutrons will do so.
Such a situation may  have many  implications  for different observations  of neutron stars which we discuss.
 \end{abstract}

\section{Introduction}

One of the exciting recent results in the studies of short-range  correlations~(SRCs) in  nuclei is the observation of the 
strong  (by factor of 20) dominance  of the  $pn$ SRCs,  relative to the   $pp$ and $nn$ 
correlations, for nuclear internal momenta  of $\sim 300-600$~MeV/c\cite{isosrc,EIPsc}. 

This observation is understood\cite{isosrc, t2,Sch} 
based on the dominance of the tensor forces in NN interaction at the above mentioned momenta  corresponding to 
average nucleon separations of $\sim 1$~Fm.  At these distances the dominating NN central  potential crosses  the 
zero due to transition from attractive to repulsive interaction allowing  tensor forces to dominate in  this transition range.
The tensor interaction projects the NN SRC part of the 
wave function into the  isosinglet - relative angular momentum, $L=2$  state,  almost identical to the $D$-wave component of 
the deuteron wave function.  As a results $pp$ and $nn$ components of the NN SRC will be strongly suppressed since they 
are dominated by the central NN potential with relative $L=0$. 
The {\em resulting picture} for the nuclear matter consisting of 
protons and neutrons  at densities  in which inter-nucleon distances are $\sim 1$~Fm is rather unique:  
it represents a system  with  suppressed  $pp$ and $nn$   but  enhanced   $pn$  interactions.

The goal of our study is to understand the implication of the above described conditions on the 
the  momentum  distribution of protons and neutrons in high density nuclei matter.

\section{New  Relation between  High Momentum $p$- and $n$-distributions in Nuclei:} 

Due to short range nature of $NN$ interaction the  nuclear  momentum distribution, $n^A({p})$,  for momenta, $p$,  
exceeding  the characteristic nuclear Fermi momentum $k_{F}$ is predominantly defined by the momentum distribution 
in the SRCs.  There is a rather large experimental body of information indicating that for the range of $k_F < p \le 600$ MeV/c 
the SRCs are dominated by 2N correlations, which  consist of mainly  the $pn$  pairs (for recent reviews see \cite{srcrev,srcprogress}).  

In recent work\cite{newprops}  based on the dominance of the $pn$ SRCs  we predicted two new properties for the nuclear 
momentum distributions at $\sim k_F < p < 600$: \\
(i) There is an approximate equality of $p$- and $n$- momentum distributions 
weighted by their relative fractions in the nucleus $x_p = {Z\over A}$ and $x_{n} = {A-Z\over A}$:
\begin{equation}
x_p n^{A}_{p}({p}) \approx x_n n^A_n({p}),
\label{p=n}
\end{equation}
with $\int n^{A}_{p/n}({p})d^3p = 1$. \\
(ii) The probability of proton or neutron being in high momentum NN SRC is inverse proportional to their relative fractions and 
can be related to the momentum distribution in the deuteron $n_d({p})$ as:
\begin{equation}
 n^{A}_{p/n}({p}) \approx {1\over 2 x_{p/n}} a_2(A,y)\cdot n_d({p}),
 \label{highn}
 \end{equation}
where $a_2(A,y)$ is interpreted as a per nucleon probability 
of finding 2N SRC in the given $A$ nucleus\cite{FS81,FS88,FSDS}  
and the nuclear asymmetry parameter  is defined as  $y= |1- 2 x_p| = |x_n - x_p|$.  

The above two properties are obtained assuming no  contributions from $pp$, $nn$ 
as well as higher order SRCs.  They follow from the assumption that the  whole  
strength  of nuclear  high momentum distribution as well as per nucleon  probability of proton and neutron to 
be in the SRC is defined by the  {\em same}  $pn$ correlation. 

These  properties can be checked for light nuclei by direct calculations  of momentum distribution of nucleons 
in asymmetric nuclei (see below).

Predictions can be made also for heavy nuclei and infinite nuclear matter if one estimates the $a_2$ parameters entering 
in Eq.(\ref{highn}) for large $A$ and extrapolate for infinite nuclear matter. 
Since  SRCs are defined by local properties of nuclei, one expects that the $A$ dependence of $a_2$   is related 
to  the nuclear density, i.e. $a_2(A,y) = a_2(\rho,y)$.  This could   allow  us to evaluate the high 
momentum part of the nucleon momentum distribution  not only for finite\cite{newprops}  but also for infinite 
nuclear matter.\\

\section{High Momentum Features of Light Nuclei} 
One can check the validity of 
the above two (Eqs.(\ref{p=n}) and (\ref{highn})) observations for light nuclei for which  it is possible to 
perform realistic calculations  based on the Faddeev equations for A=3 systems\cite{Bochum},
Correlated Gaussian Basis(CGB) approach\cite{CGB}  as well as  Variational Monte Carlo method(VMC)
for light nuclei  A (recently being available for up to $A =12$ \cite{VMCpc}).

\begin{figure}[ht]
\vspace{-0.4cm}
\centering\includegraphics[scale=0.45]{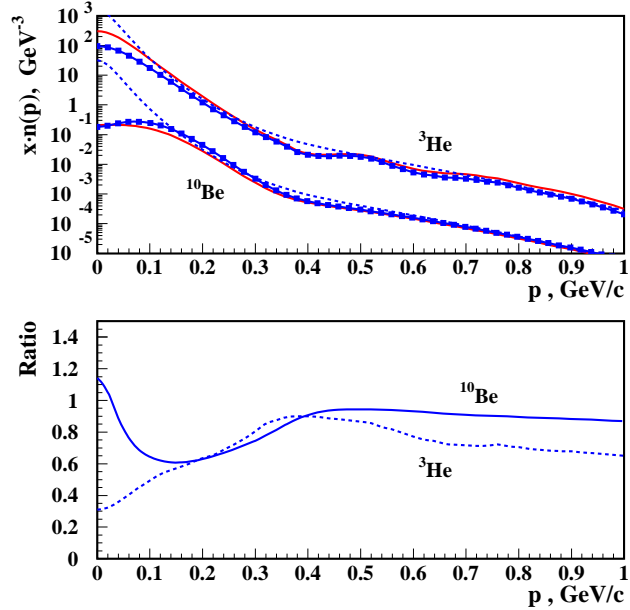}
\vspace{-0.4cm}
\caption{(a) The momentum distributions of proton  and neutron weighted by $x_p$ and $x_n$ respectively.
The doted lines represent the prediction for the  momentum distribution according to Eq.(\ref{highn}).  \ 
 (b)  The $x_{p/n}$ weighted ratio 
of  neutron to  proton  momentum distributions.  See the text for details.}
\label{He3_Be10}
\end{figure}

The validity of Eq.(\ref{p=n}) is checked in Fig.\ref{He3_Be10} for 
$^3He$ nucleus,  based on the solution of  Faddeev equation\cite{Bochum},   and 
for $^{10}Be$  based on VMC calculations\cite{VMCpc}.  
The solid  lines with and without squares in Fig.\ref{He3_Be10}(a)  represent 
neutron and proton momentum distributions for both nuclei weighted by their respective 
${x_n}$ and ${x_p}$ factors.

As one can see for $^3He$  the proton momentum distribution 
dominates the neutron  momentum distribution at small momenta reflecting the fact that in the mean field 
the probability of finding proton is larger than neutron   just because there are twice as much protons in $^3He$ than 
neutrons.
The same is true for $^{10}Be$ for which now the neutron momentum distribution dominates at small momenta.
However   at  $\ge 300$~MeV/c  for both nuclei, the proton and neutron momentum distributions become close  to each other 
up to the  internal momenta of $600$~MeV/c. This is the region dominated by tensor interaction. As Fig.\ref{He3_Be10}(a) 
shows the high momentum distribution modeled according to Eq.(\ref{highn}) (dotted lines) agrees reasonably well  with the 
realistic calculations.

This effect is more visible for the ratios  of weighted  n-  to p- momentum 
distributions in Fig.\ref{He3_Be10}(b), demonstrating  that the approximation of Eq.(\ref{p=n}) in the range of 
$300-600$ MeV/c  is good  on the level of 15\%.     Note that the similar features  present for 
all other asymmetric nuclei calculated  within the VMC method in Ref.\cite{VMCpc}.

The  prediction of Eq.(\ref{highn}) can be checked by comparing the kinetic energies of  proton and neutron, in which case we 
expect that  per nucleon kinetic energy of the lesser component to be larger.  As  it can be seen from Table~\ref{table2} this prediction is  confirmed too by realistic calculations.

\begin{table}[t]
\vspace{-0.4cm}
\caption{Kinetic energies (in MeV) of proton and neutron} 
\medskip
\centering 
\begin{tabular}{l l l l c }
\hline\hline 
A & y & $E_{kin}^p$  &   $E_{kin}^n$  &   $E_{kin}^p-E_{kin}^n$\\  [0.0ex] 
\hline 
$^{8}$He     \    & 0.50  \  & 30.13 \  &  18.60 \  & 11.53     \\ 
$^{6}$He     \    & 0.33  \  & 27.66 \  & 19.06  \  &  8.60  \\ 
$^{9}$Li       \    & 0.33  \  & 31.39 \  &  24.91 \  & 6.48    \\ 
$^{3}$He     \    & 0.33  \  & 14.71 \  & 19.35  \  &  -4.64   \\ 
$^{3}$He\cite{Bochum} \ & 0.33 \ & 13.70 \ & 18.40 \ &  -4.7 \\
$^{3}$He\cite{CGB} \ & 0.33 \ & 13.97 \ & 18.74 \ &  -4.8 \\
$^{3}$H       \    & 0.33  \  & 19.61   \  &  14.96 \  &   4.65   \\ 
$^{8}$Li       \    & 0.25  \  & 28.95 \  &  23.98 \  & 4.97   \\ 
$^{10}$Be   \    & 0.2    \   & 30.20 \  &  25.95 \  & 4.25    \\
$^{7}$Li       \    & 0.14  \  & 26.88 \  &  24.54 \  & 2.34   \\ 
$^{9}$Be     \    & 0.11  \   & 29.82 \  &  27.09 \  & 2.73    \\
$^{11}$B     \    & 0.09   \  \  \ & 33.40 \  \  \ &  31.75 \ \ \  &   1.65    \\ 
 \hline 
\end{tabular}
\label{table2} 
\end{table}

\section {High Momentum Properties of Heavy Nuclei}
Presently, no realistic calculations exist for asymmetric heavy nuclei for the predictions of Eqs.(\ref{p=n}) and (\ref{highn}) 
to be checked.    The main   predictions of  Eq.(\ref{highn}) is that 
high momentum   protons and neutrons became  increasingly  {\em unbalanced}  with an  increase 
of the nuclear  asymmetry,  $y$. To quantify this, using Eq.(\ref{highn})  one can calculated the fraction 
of the   nucleons having  momenta $\ge k_{F}$ as:
\begin{equation}
P_{p/n}(A,y) \approx  {1\over 2 x_{p/n}} a_2(A,y)\int\limits_{k_F}^{\infty} n_d(p) d^3p,
\label{fraction}
\end{equation}
where the parameter $a_2(A,y)$ can be taken from the experimental analysis of inclusive $A(e,e^\prime)$  data at 
kinematics dominated by SRC\cite{FSDS,Kim1,Kim2,Fomina2}.   
The results of the estimates of $P_{p/n}(A,y)$ for medium to heavy nuclei are presented in Table 2.
 
\begin{table}[th]
\vspace{-0.4cm}
\caption{Fractions of high momentum ($\ge k_F$) protons and neutrons in nuclei A}
\medskip
\centering
\begin{tabular}{llllll}
\hline\hline 
A & $P_p(\%)$  & $P_n(\%)$ & A & $P_p(\%)$  & $P_n(\%)$ \\  [0.5ex] 
 \hline
12  & 20 & 20 &
56  & 27 & 23 \\
27  & 24 & 22 &
197 & 31 & 20 \\
\hline
\end{tabular}
\label{table1}
\end{table}
As it  follows from the table, as  the asymmetry increases the imbalance between the high momentum 
fractions of proton and neutron grows. 
For example,   in the Gold,  the relative fraction of high momentum ($\ge k_{F}$)  protons  is  50\% more  
than that of the neutrons. 

Such a dominance of the proton high momentum component in heavy neutron rich nuclei can be checked in 
high momentum transfer single proton knock-out processes in which long-range nuclear effects are well controlled\cite{gea,ms01}.  
The first such experimental  verification 
for heavy nuclei  is currently   underway in quasi elastic $A(e,e,p)X$ measurement at Jefferson Lab, where  
the ratio of  high momentum fractions of  nucleons in $^{56}Fe$ and $^{208}Pb$  to that of $^{12}C$ is extracted.  The
results\cite{EIPpc} are in reasonably good agreement with the prediction of  Eq.(\ref{fraction}) (Table 2)  and 
they are being prepared for publication.  

\section {High Momentum Properties of Asymmetric Nuclear Matter}
To estimate the fractions of energetic protons and neutrons in asymmetric nuclear matter using Eq.(\ref{fraction})  one needs to 
extrapolate $a_2(A,y)$ for infinite nuclear matter.  This was achieved in Ref.\cite{proa2} where based on the local property of 
SRCs the  $a_2(A,y)$ parameters were represented as a function of local nuclear density $\rho$ and asymmetry $y$: $a_2(\rho,y)$.  Then 
all the available data  on inclusive $A(e,e')X$ scattering\cite{DDay} at SRC kinematics were used to fit and extrapolate $a_2$ magnitudes for 
saturation and above saturation densities of nuclear matter at given asymmetry $y$. The accuracy of the fit was checked for symmetric nuclear matter 
at the saturation density,  for which the obtained value of $a_2(\rho_0,0) \approx 7.03\pm 0.41$ is in reasonable 
agreement with other  estimates of $a_2$ for symmetric nuclear matter\cite{CPS}.

For the case of asymmetric nuclear matter we considered a neutron star matter, 
in which for asymmetry parameter  $y$,  we used the threshold value of 
$x_p = {1\over 9}$ ($y={7\over 9}$) below of which the direct URCA processes:
\begin{equation}
n\rightarrow p + e^- + \bar \nu_e, \ \ \ \ p+ e^-\rightarrow n+\nu_e
\label{DURCA}
\end{equation}
will stop in the standard model of superdense nuclear matter consisting of degenerate 
protons and neutrons\cite{LPPH}.   Estimating the Fermi momenta of protons and 
neutrons in Eq.(\ref{fraction}) with $k_{F,N} = (3\pi^2x_N\rho)^{1\over 3}$, in Fig.2 we 
present the off-Fermi-shell fractions of protons and neutrons as a function of nuclear density.
The most interesting result of these estimates is that in the equilibrium,  $pn$  SRCs
move the large fraction of protons above the Fermi-shell:  at 
$3\rho_0$ densities  half of the protons will be  off-Fermi-shell while at $\rho \ge 4.5\rho_0$ 
all the protons will populate the high momentum tail of the momentum distribution.  
The situation however is not as dramatic for neutrons,  with only  about 2~\% of neutrons 
populating the high momentum part of the momentum distribution.\\

\begin{figure}[ht]
\vspace{-0.4cm}
\centering\includegraphics[height=6cm,width=9cm]{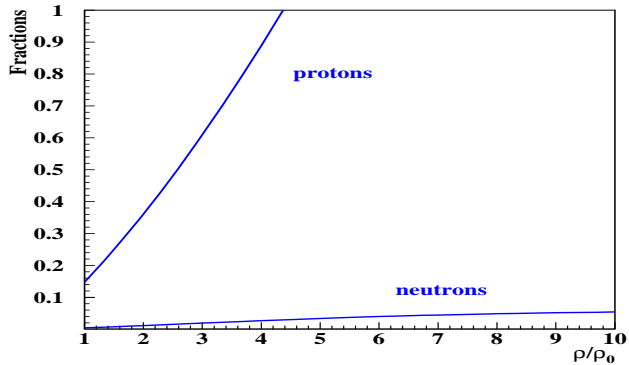}
\vspace{-0.4cm}
\caption{Density dependence of the fraction of off-Fermi-shell nucleons in $x_p = {1\over 9}$ matter.}
\label{QIM}
\end{figure}

It is worth mentioning that the present result of strong modification of proton momentum distribution in high density 
asymmetric matter  is in qualitative agreement with the nuclear matter calculation  based on Green function method\cite{Rios}.

\section{Possible Implications for  Nuclei and  Neutron Stars:}
Our main observation is that with an increase of nuclear asymmetry the lesser  component  become more energetic. 
This is confirmed\cite{newprops} for light nuclei by direct estimates  of the average kinetic energies of 
proton and neutron  using realistic wave function calculations (Table 1).
In the case of  neutron rich heavy nuclei ($A\ge 40$),  a larger fraction of protons will occupy the high momentum 
tail of the momentum distributions (Table 2). 
This may have several verifiable implications for large $A$  nuclear phenomena\cite{newprops}. Particularly our 
estimates show that it may explain\cite{memc} the large A part of the recently observed correlation between the strengths of the 
EMC (nuclear partonic distribution modification)  and SRC  effects\cite{Larry1}.   

The energetic protons in neutron rich nuclei will result  also to the stronger  nuclear modification of 
$u$-quarks  as compared  to $d$-quarks  and the effect will  grow with  A. This 
provides  an explanation\cite{aps} of the NuTeV anomaly\cite{Zeller}.  The predicted effect also 
can be  checked in parity violating deep inelastic scattering off heavy nuclei.

However, our observation may have  more dramatic  implications for  the dynamics of  neutron stars. Some of them are:\\
- {\em Cooling of a Neutron Star:} Large concentration of protons above the Fermi momentum 
will allow the condition for Direct URCA processes  $p_p + p_e > p_n$\cite{LPPH} to be satisfied 
even if $x_p < {1\over 9}$.  This will allow a  situation in which intensive cooling of 
the neutron stars continues well beyond the critical point $x_p = {1\over 9}$
(see also Ref.\cite{srcrev}).

\noindent
- {\em Superfluidity of Protons:} Transition of  protons to the high  momentum tail 
will smear out the energy gap which will remove the superfluidity condition for the protons.  

\noindent
- {\em Protons in the Neutron Star Cores:} The concentration of  protons in the high momentum 
tail will result in  proton densities $\rho_p\sim p_p^3\gg k_{F,p}^3$.  This   will favor 
an equilibrium condition with "neutron skin" effect in which
large concentration  of protons populate the core rather 
than   the crust of the neutron star.  This and the proton superfluidity condition violation
may provide different dynamical picture  for generation of  magnetic fields in the stars. 
  
\noindent
- {\em Isospin locking and the stiff equation of state of  neutron stars:} With an increase 
in density  more and more protons move to the high momentum tail where they are in 
short range tensor correlations with neutrons. In this case  one would expect that 
high density nuclear matter to be dominated by configurations with quantum 
numbers of tensor correlations ($S=1,I=0$).  In such a scenario protons and neutrons at 
large densities will be locked in the NN iso-singlet state.  This will 
double the threshold of inelastic excitation from $NN\rightarrow N\Delta$ to 
$NN\rightarrow \Delta\Delta (NN^*)$ transition thereby stiffening the equation of  state which is 
favored by the recent  large neutron star mass observation\cite{NSM}.\\

\section{Possible Universality of the Obtained Results}  
Our observation is relevant to any 
asymmetric two-component Fermi system in which the interaction within  each component is suppressed 
while the mutual interaction between two components is enhanced. It is interesting that the similar situation 
is realized for two-fermi-component ultra-cold atomic systems\cite{Shin:2006zz}  but with the mutual s-state 
interaction.
One of the most intriguing aspects of   such systems is that in the asymmetric limit  they exhibit very rich 
phase structure with indication of the  strong modification of the small component of the mixture\cite{Bulgac1}. In this 
respect our case is similar to that of ultra-cold atomic systems with the difference that the 
interaction between components has a tensor nature.\\

\noindent
This work is supported by U.S. DOE  grant under contract DE-FG02-01ER41172.


\section*{References}
\begin{thebibliography}{9}

\bibitem{isosrc} E.~Piasetzky, M.~Sargsian, L.~Frankfurt, M.~Strikman 
and J.~W.~Watson, Phys.\ Rev.\ Lett.\  {\bf 97}, 162504 (2006).

\bibitem{EIPsc}R.~Subedi {\it et al.},  Science {\bf 320}, 1476 (2008).

\bibitem{t2}
  M.~M.~Sargsian, T.~V.~Abrahamyan, M.~I.~Strikman and L.~L.~Frankfurt,
   Phys.\ Rev.\  C {\bf 71}, 044615 (2005).

\bibitem{Sch}R.~Schiavilla, R.~B.~Wiringa, S.~C.~Pieper and J.~Carlson,
  Phys.\ Rev.\ Lett.\  {\bf 98}, 132501 (2007).

        
\bibitem{srcrev}L.~Frankfurt, M.~Sargsian and M.~Strikman,
  Int.\ J.\ Mod.\ Phys.\  A {\bf 23}, 2991 (2008).
  
\bibitem{srcprogress}  J.~Arrington, D.~W.~Higinbotham, G.~Rosner, M.~Sargsian,
  Prog.\ Part.\ Nucl.\ Phys.\  {\bf 67}, 898 (2012).
  
 \bibitem{newprops}  M.~M.~Sargsian,  arXiv:1210.3280 [nucl-th].
 
  \bibitem{FS81}L.~L.~Frankfurt and M.~I.~Strikman,
        Phys.\ Rept.\  {\bf 76}, 215 (1981).


\bibitem{FS88}L.~L.~Frankfurt and M.~I.~Strikman,   Phys.\ Rept.\  {\bf 160}, 235 (1988).

\bibitem{FSDS}L.~L.~Frankfurt, M.~I.~Strikman, D.~B.~Day and M.~M.~Sargsian,
  Phys.\ Rev.\  C {\bf 48}, 2451 (1993).


 
\bibitem{Bochum}A.~Nogga, H.~Kamada and W.~Glockle, 
                  Nucl.\ Phys.\ {\bf A689}, 357 (2001).
\bibitem{CGB}  H.~Feldmeier, W.~Horiuchi, T.~Neff and Y.~Suzuki,
  Phys.\ Rev.\ C {\bf 84}, 054003 (2011).

\bibitem{VMCpc}   R.~B.~Wiringa, R.~Schiavilla, S.~C.~Pieper and J.~Carlson,
  arXiv:1309.3794 [nucl-th].

\bibitem{Kim1}K.~S.~Egiyan {\it et al.}, 
  Phys.\ Rev.\  C {\bf 68}, 014313 (2003).

\bibitem{Kim2}K.~S.~Egiyan {\it et al.}, 
        Phys.\ Rev.\ Lett.\  {\bf 96}, 082501 (2006).


\bibitem{Fomina2}   N.~Fomin {\it et al.},  Phys.\ Rev.\ Lett.\  {\bf 108}, 092502 (2012).



 
\bibitem{gea} 
  L.~L.~Frankfurt, M.~M.~Sargsian and M.~I.~Strikman,
  Phys.\ Rev.\ C {\bf 56}, 1124 (1997).
   
\bibitem{ms01} M.~M.~Sargsian, Int.\ J.\ Mod.\ Phys.\ E {\bf 10}, 405 (2001).
  
\bibitem{EIPpc}E. Piasetzky, 
http://www.int.washington.edu/talks\\/WorkShops/int\_13\_52W/People/Piasetzky\_E/Piasetzky.pdf.


\bibitem{proa2} M.~McGauley, M.~M.~Sargsian, arXiv:1102.3973 [nucl-th].

\bibitem{DDay}O.Benhar, D.Day and I.Sick,  [arXiv:1104.1196 [nucl-ex]].

\bibitem{CPS}C.Ciofi degli Atti, E. Pace, G.Salme, Phys.\ Rev.\ C {\bf 43}, 1155 (1991).

\bibitem{LPPH} J.~M.~Lattimer, M.~Prakash, C.~J.~Pethick and P.~Haensel,
         Phys.\ Rev.\ Lett.\  {\bf 66}, 2701 (1991).

\bibitem{NSM} P.~Demorest, T.~Pennucci, S.~Ransom, M.~Roberts and J.~Hessels,
        Nature {\bf 467}, 1081 (2010).
\bibitem{Rios} A.~Rios, A.~Polls and W.~H.~Dickhoff,   Phys.\ Rev.\ C {\bf 79}, 064308 (2009).


\bibitem{memc}   M.~M. Sargsian,  arXiv:1209.2477 [nucl-th].

\bibitem{Larry1}   L.~B.~Weinstein, {\it et al.},
  Phys.\ Rev.\ Lett.\  {\bf 106}, 052301 (2011).
   

\bibitem{aps}M.~Sargsian, "Neutrino interactions in the nuclear environment", APS Meeting, 
April, 2013; Denver, Co.

\bibitem{Zeller}   G.~P.~Zeller {\it et al.}  [NuTeV Collaboration],
  Phys.\ Rev.\ Lett.\  {\bf 88}, 091802 (2002).

\bibitem{Shin:2006zz}   Y.~Shin, M.~W.~Zwierlein, C.~H.~Schunck, A.~Schirotzek and W.~Ketterle,
  Phys.\ Rev.\ Lett.\  {\bf 97}, 030401 (2006).       
\bibitem{Bulgac1} A.~Bulgac and M.~M.~Forbes,   Phys.\ Rev.\ A {\bf 75}, 031605 (2007).


 
\end{thebibliography}
\end{document}